# Cyber Security Standards, Practices and Industrial Applications:
Systems and Methodologies

# Embedded Systems Security

**Muhammad Farooq-i-Azam**
*COMSATS Institute of Information Technology, Pakistan*

**Muhammad Naeem Ayyaz**
*University of Engineering and Technology, Pakistan*

## ABSTRACT

*Not long ago, it was thought that only software applications and general purpose digital systems i.e. computers were prone to various types of attacks against their security. The underlying hardware, hardware implementations of these software applications, embedded systems, and hardware devices were considered to be secure and out of reach of these attacks. However, during the previous few years, it has been demonstrated that novel attacks against the hardware and embedded systems can also be mounted. Not only viruses, but worms and Trojan horses have been developed for them, and they have also been demonstrated to be effective. Whereas a lot of research has already been done in the area of security of general purpose computers and software applications, hardware and embedded systems security is a relatively new and emerging area of research. This chapter provides details of various types of existing attacks against hardware devices and embedded systems, analyzes existing design methodologies for their vulnerability to new types of attacks, and along the way describes solutions and countermeasures against them for the design and development of secure systems.*

## INTRODUCTION

A few years ago almost all electronic equipment was built using analog components and devices. However, after the advent of microprocessors and microcontrollers majority of electronic equipment developed today uses digital components for design implementation. Embedded systems are finding their use in diverse applications ranging from complicated defense systems to home gadgets. Smart cards, debit and credit cards, DVD players, cell phones and PDAs are just a few examples of embedded systems that we use in our daily lives.

Under certain circumstances and conditions, a larger digital system is usually dependent upon the functions of smaller component embedded systems for its function and operation. For example, a general purpose computer houses many smaller embedded systems. A hard disk, a network interface card, CD-ROM drive are examples of embedded systems used by a computer system for its operation. In addition to this, large industrial plants, nuclear power plants, passenger and fighter aircrafts, weapons systems, etc. are a few of many places where embedded systems are part of a bigger system.

With this increased usage of embedded systems in our daily lives, it is not unusual that bad guys and criminals try to take advantage of weak links in their security. Specially, the embedded systems used in financial institutions, battlefield equipment, fighter planes and industrial and nuclear plants may become targets of attack due to the importance of functions performed by them. Therefore, it is essential that these systems and the components used in them are highly dependable and their security is not compromised.

A number of security incidents related to embedded systems have been reported in the literature. For example, in 2001, Shipley and Garfinkel found an unprotected modem line to a computer system which was being used to control a high voltage power transmission line (Koopman, 2004). In another incident, a disgruntled employee in Australia released almost 250 million tons of raw sewage by causing failure of control system of a waste treatment plant through a remote attack (IET, 2005).

It is pertinent to mention here that the organizations which become target of attack may not like to publicize the incident due to various reasons. For example, it may disclose a vulnerable area of their systems or it may cause them a bad name and raise questions against security of their other assets. Furthermore, security threats against embedded systems do not propagate as rapidly as those against a standard operating system or software application. This is because majority of personal computer systems is similar and it is easier for any security threat to replicate from one system to the other. On the other hand, each embedded device is unique and it is almost impossible for a security threat to propagate from one device to the other. Moreover, a security threat against an embedded device is generally initiated at any one of the design stages before the device is built. Security threats against a software system may be programmed at any time after they have been developed and deployed. These are a few of the many reasons that we do not come to see as many security incidents reported against embedded systems as against software applications. Despite this fact, security incidents have been reported against hardware devices and embedded systems, a couple of which have been cited above and a few more will be mentioned later in this chapter.

## BACKGROUND

Embedded systems security is a new and emerging area of research. It is meeting point of many disciplines such as electronics, logic design, embedded systems, signal processing and cryptography. It is closely related to the area of information and software systems security because software is an integral component of any embedded system.

First microprocessor was developed around 1971 and later innovations in this field resulted in the development of computer systems and embedded devices. Software is an integral component of the both. In particular, every desktop computer carries a critical piece of software called the operating system. It manages the hardware resources and makes it possible for an end user to operate the computer. Other software applications in a computer run on top of the operating system.

It was the software component of digital systems which was first subjected to different types of security threats and attacks and many security incidents were reported against different operating systems and software applications. This started in 1970s and continues to date. However, embedded systems security gained importance in 1990s, specially, after side channel attacks were shown to be successful against smart cards. Later, emergence of networked embedded systems highlighted this area of research as the embedded devices could now be subjected to remote attacks.

Many of the methods and techniques used in the attacks against software applications can also be used against embedded devices, specially, in the firmware component. However, a few considerations involving the security of an embedded system are different from those of a general purpose digital system. To get a better perspective, it would help to look at the traits of embedded systems security which are different from those of software security.

### Embedded Systems Security Parameters

An embedded system is a digital device that does a specific focused job as compared to a general purpose digital system such as a personal computer. Whereas a general purpose digital system can be used for a variety of tasks by installing new software, the software for an embedded system is generally fixed and has limited flexibility in allowing user programs to run. For example, an operating system in a desktop computer allows user to perform a variety of tasks by installing appropriate software. The software can later be un-installed, modified or updated without much of a hassle. However, in the case of an embedded system, this is not the case. A traffic light controller, for example, is a dedicated system performing a specific function. The software in such a digital system has limited flexibility and usually does not allow user to install new software on top of the base software. Also modification and up-gradation of software is not as easy as in the case of desktop computer. The software for such a system normally resides in Electrically Erasable and Programmable Read Only Memory (EEPROM) which has to be re-programmed using EEPROM programmers. In the context of security, it has an important implication i.e. if software in an embedded system is compromised, it will be lot more difficult to replace or upgrade as compared to software in a general purpose digital system. There are usually no software upgrades and patches for bugs as far as embedded devices are concerned.

Embedded systems have limited resources e.g. small memory, no secondary storage device and small input and output devices. These limitations provide an avenue of attack in that a software virus or a hardware Trojan horse can cause denial of service by consuming any of these resources. For example, many of the embedded systems have energy constraints and are often battery powered. These may have to operate over battery over an extended period of time and the power consumption has to be very low. By seeking to drain the battery, an attacker can cause the system failure even when breaking into the system is impossible.

Embedded systems generally carry out tasks that have timing deadlines. Missing these timing deadlines may cause loss of property or even life. Again this is a unique attack vector against embedded systems. By simply, adding some sort of delay in the execution of an instruction or a series of instructions, the attacker can achieve the objective of the attack.

While desktop systems and other similar equipment may operate in an environment where temperature and environment conditions are controlled to meet the requirements of the equipment installed, some of the embedded systems usually operate under extreme environment conditions, e.g. high temperature, humidity and even radiation. Causing any of these environment parameters to change can also affect the performance of embedded systems.

General purpose computers usually employ a popular brand of processors. Whereas, in the case of embedded systems there is a lot of variation in processors and operating systems. This provides for an inherent security against the propagation of attacks and security threats from one device to the other.

Embedded systems might operate unattended without the need of a system administrator. Therefore, there is usually no reporting mechanism on the attacks being carried out against the device similar to those reported by an antivirus or firewall in the case of traditional computer systems.

Construction of embedded systems invariably depends upon hardware devices and components such as Integrated Circuits (ICs). There is a broad array of attacks which can be mounted against an embedded system either at the level of firmware or at the level of circuit. Further, circuit level attacks may exist either at the level of discrete components or hidden in an integrated circuit. In other words, security issues in an embedded system do not confine themselves to a single layer of abstraction but rather span across various layers of abstraction from tiny hardware components to firmware and software. In addition, the discrete components or the integrated circuit may or may not be part of an embedded system.

## SECURITY ISSUES IN EMBEDDED SYSTEMS

A system is secure if it could be used only for the purpose for which it is intended and only by the prescribed and authorized user and is available to provide service at any time. This statement is also true for embedded systems in general.

Before we discuss the types of attacks and security issues, it is imperative to understand the lifecycle and design and development methodologies related to hardware devices and embedded systems.

Hardware devices and embedded systems can be implemented in a number of ways depending upon the application of the particular system under development. For example, we can develop a typical embedded system using an appropriate microcontroller and peripheral components. Similarly, for some other embedded system, an implementation using a Field Programmable Gate Array (FPGA) may be more appropriate. Whatever strategy for implementation of the embedded system we may choose, there are various stages involved from design to implementation. Figure 1 below shows various stages or layers of abstraction from design to implementation for an embedded system using an FPGA.

## Figure 1. Layers of abstraction in an FPGA based design.

A security loophole may be present at any of these layers and detection becomes more and more difficult as we move from top to bottom layer. This fact is shown with increasing level of gray from the top layer to the bottom layer.

**Types of Attacks**

Attacks on embedded systems can be broadly categorized as:
- Design and algorithmic attacks
- Side channel attacks

As the name suggests, a design and algorithmic attack exploits weakness inherent in the design and algorithm of the embedded system whereas a side channel attack attempts to exploit weakness in the implementation of the design. It is pertinent to point out that the bug may be left un-intentionally which is seldom the case or intentionally by the designer(s) involved at various stages of the implementation of the design.

Design and algorithmic attacks rely on the bug left in the device during the design stage in any of the layers of abstraction of the embedded system. For example, in the case of an FPGA based system, the bug may be left in the program code of the system which is usually written using a Hardware Description Language (HDL). It may also be planted intentionally by the designer after the HDL code synthesis at the logic gate level. Such a bug exists in the embedded system in the shape of tangible hardware and is commonly known as hardware Trojan horse. In a similar fashion, the bug may be planted at the transistor level design or indeed at the level of semiconductor in the case of an integrated circuit. In the case of a microcontroller and microprocessor based embedded system, the bug may be planted in the control program of the microcontroller. Though the bug exists in the software code, it may still be called hardware Trojan horse as it alters the designated behavior of the embedded system to achieve hidden objectives.

In the case of side channel attacks, the attacker treats the embedded system as a black box and analyzes the inside of the system by feeding it various types of inputs and then observing the behavior of the system and its output. These types of attacks are normally used to extract some secret information from the embedded system.

**HARDWARE TROJAN HORSES**

Consider the case of standalone embedded systems i.e. an embedded system that is not part of any network. As the embedded device does not interact with any external network, it may be thought that no attacks can be mounted against the device. However, it is still possible for a malicious design engineer to leave a malignant hole i.e. a Trojan horse in the system. For example, a design engineer could program an embedded device to run correctly for all operations except, say, #2600th, or program it in such a way to behave erratically after certain number of operations or under a certain critical condition. If the device is part of a critical safety system in, say, an industrial process plant, the consequences may have a devastating effect on the plant operation. The result may be degraded performance, partial shutdown of the process or even complete failure of the entire plant.

The designer could also leave a hardware Trojan horse that can be controlled remotely, possibly using a radio channel, in an otherwise standalone embedded system. It may be noted that the Trojan horse may be part of an embedded system in the shape of discrete components or at the level of transistors in an integrated circuit. In the first case, the Trojan horse may be present on a circuit board of the embedded system, and in the second case, it may be present inside an integrated circuit in the shape of transistors.

It would be interesting to cite just two instances where hardware Trojan horses were detected and found spying on unauthorized information. In the first instance, Seagate external hard drives were found to have a hardware Trojan horse that transmitted user information to a remote entity (Farrell, 2007). In the second instance, Prevelakis and Spinellis (2007) report that, in 2006, Vodafone routers installed in Greece were altered in such a manner so as to allow eavesdropping phone conversation of the prime minister and many other officials.

As discussed earlier, the hardware Trojan horse may be implanted in the hardware device or embedded system at any of the various stages from design to implementation including plantation. This includes the possibility of implant by the designer at the level of behavioral description, by a third party synthesis tool or by the fabrication facility.

Alkabani and Koushanfar (2008) describe a method by which a designer could leave a Trojan horse in the pre-synthesis stage which after going through the rest of stages of implementation becomes part of the circuit. The designer first completes a high level description of the system to arrive at the Finite State Machine (FSM) of the design. It is at this stage that the designer manipulates the FSM by introducing more states into it such that the state transitions and their inputs can be controlled by a hidden input which may be used to trigger a hidden function in the system. As FSM is typically used to specify control part of the design, it occupies a very small fraction to the scale of only 1% of the total area and power consumption of the complete design. Therefore, even if the designer wishes to insert a large number of states which is two or three times of the original number, it is still only a small part of the complete system allowing it be hidden from detection.

Potkonjak, Nahapetian, Nelson and Massey (2009) in their paper describe a hardware Trojan horse meant for a cell phone. The Trojan triggers on the reception of a call from a certain caller ID. The Trojan can either conference call a third party so as to leak all the conversation or it can even render the phone useless.

Clark, Leblanc and Knight (in press) describe a novel hardware Trojan horse which exploits unintended channels in the Universal Serial Bus (USB) protocol to form a covert channel with a device to which it is connected. USB protocol is used to interface a multitude of devices to embedded and computer systems. Keyboard, mouse and speakers are a few of the many devices which use USB interface to connect to a digital system. USB uses two communication channels to interface a keyboard to a digital system. One channel, which is the data channel and is unidirectional, transmits key strokes to the digital systems and the other channel, which is a control channel and is bidirectional, transmits and receives control information, e.g. CAPS lock, NUM lock and SCROLL lock keys to and from the digital system. These are the intended uses of the channels and these channels are not meant to be used for the transmission and reception of other information. Similarly, USB allows interfacing audio speakers to a digital system using control and data channels like those used for keyboard. J. Clark et al. modified the standard use of these channels so as to form a covert channel to transmit and receive information other than meant to be communicated by these channels. Using these primitives, a standard USB keyboard can house additional components so as to also function as a hardware Trojan horse that utilizes the covert channels described above. If such a keyboard is connected to a digital system such as a computer, the security system at the computer will not object to the legitimacy of the USB device as it presents itself as a keyboard to the computer system. After it interfaces to the computer system, it can behave as a standard keyboard and in addition also log sensitive information such as username and password of an important resource such as the computer system itself. Using the username and password, the USB-based hardware Trojan horse can then upload malicious software application such as a software Trojan horse at any time of its convenience. An illustrative representation of such a hardware Trojan horse employing covert USB channels is shown in the Figure 2.

Figure 2. USB-based Hardware Trojan Horse Using Covert Channels

While a few hardware Trojan horse implementation mechanisms have been suggested, examples of which are given above, some detection mechanisms for the hardware Trojan horses have also been proposed. It may be noted that complexity of design and implementation mechanisms adds to the difficulty of detecting a hardware Trojan horse.

Modern hardware devices and circuits contain a large number of gates, transistors, I/O pins. In addition to this, there is a large variety of components in a single hardware device or integrated circuit and a hardware Trojan horse can be implemented in a number of ways. All this added together makes the detection of hardware Trojan horse difficult and specialized task.

Potkonjak, Nahapetian, Nelson, and Massey (2009) have proposed a hardware Trojan horse detection mechanism for an integrated circuit using gate-level characterization. The technique makes non-destructive measurements of certain characteristics of gates in the Integrated Circuits (ICs). These characteristics include timing delay, power consumption and leakage current. Temperature and radiation can also be considered for this purpose. This measurement helps approximating the scaling factor of most of the gates in the integrated circuit. Further, a programming model comprising a linear system of equations is constructed and a statistical analysis helps determine the presence and even location of the hardware Trojan horse. However, this detection technique is limiting in cases where the added gate has the same inputs as a gate of the same type in the original design. Therefore, an attacker can circumvent this detection technique by developing a Trojan logic in which all the gates in the Trojan circuitry have the same number of inputs as similar gates in the original design which is not something difficult to achieve.

Another mechanism to detect hardware Trojan horse in an integrated circuit, proposed by Jin and Makris (2008) computes a fingerprint of path delays in the integrated circuit by performing measurements on some random samples of normal ICs. The path delays of the suspect chip are also measured and compared against the standard fingerprint. If the results do not match and variation in the path delays of the IC under test is beyond a certain threshold, the IC is marked to have a hardware Trojan horse.

While we are on the topic of hardware Trojan horse, it would be pertinent to mention BIOS virus. BIOS (Basic Input Output System) is the piece of software which resides in the motherboard Read Only Memory (ROM). A few years back, one needed to adopt elaborate procedures using EEPROM programmers to update the software in the ROM. One needed to remove the EEPROM from the motherboard, put it into an EEPROM programmer, and only then one could burn new BIOS program into it. For all this, one needed physical access to the computer system. However, with the emergence of flash ROMs, which is a particular type of EEPROM, the BIOS firmware can be updated without having it to remove from the motherboard. One can now update the firmware in the flash ROM by running a program on the computer where it is installed. It is an advantage and has made life simple for the hardware developers and end users. However, it has also made possible what we have termed as BIOS virus. Once a computer virus gains access to a computer, it can update and infect the firmware in the flash ROM as well. As a result, a remote attacker can gain access to the computer by exploiting some weakness in the application software or the operating system. Once the attacker has gained access to the computer, it can run a piece of software to change the BIOS program in the flash ROM, say, to install a BIOS level rootkit.

A BIOS virus is a piece of malicious software that infects the computer BIOS and resides in the system ROM along with regular BIOS. If a virus infects a certain application or operating system, normally a simple virus scan with antivirus software with updated virus definitions is enough to detect and remove it. However, if a virus is able to infect the BIOS, even the detection becomes difficult. For the removal also, one has to adopt special procedures apart from using regular antivirus software. This includes downloading of new BIOS and all drivers from the manufacturer's website on a clean machine, then booting the infected machine using a clean media e.g. a bootable CD/DVD ROM, porting the downloaded BIOS and drivers to the infected system using a read-only CD or USB followed by their installation on the infected machine.

## SIDE CHANNEL ATTACKS

Design and algorithmic attacks discussed above usually implant a Trojan horse in the system so that the system can perform a certain hidden action on a trigger. To be more elaborate, a hardware Trojan horse may, for example, be used to send all the records and data of the system to an unauthorized entity over a covert channel or it may be used to allow remote control of the system by an unauthorized entity. For these attacks to be effective, a certain malicious circuitry is usually part of the entire digital system, be it an embedded system based upon microcontrollers and microprocessors or be it an integrated circuit.

Side channel attacks, on the other hand, are usually used to extract some secret information stored inside a digital system. The digital system is treated as a black box and is subjected to various tests by applying different sets of stimuli to its input and noting the output behavior against every input. By comparing the output results against various inputs, an attacker tries to infer the design of the digital system and secret information stored inside it. In other words, side channel attacks exploit weakness of the implementation of the algorithm as compared to algorithmic attacks which exploit weakness in the algorithm itself.

Generally, side channel attacks are mounted against hardware implementations of various encryption algorithms so as to infer the secret key stored inside the hardware security modules. In many cases, the hardware security module under attack is a smart card which is normally used to perform encryption or decryption using the secret key stored inside it. Smart cards have found use in many applications including credit cards, payphone, GSM cell phone SIM card, etc. By seeking to extract the secret key stored inside in any of these cards, the attacker can make a duplicate or clone of the original card thus allowing him or her to use the services provided by it without the knowledge of the legitimate owner of the card. Obviously the attacker will need to have physical access to the card to perform a side channel attack on it.

Kocher (1996) is credited with the development of first side channel attack i.e. timing analysis attack and co-development of power analysis attack (Kocher, Jeffe & Jun, 1999). These attacks laid the foundation of further research leading to the development of more side channel attacks.

There are four broad categories of side channel exploits:
- Time analysis
- Error analysis
- Power analysis
- Electromagnetic Radiation Analysis

## TIME ANALYSIS

In a side channel attack based on time analysis, the attacker tries to infer protected information by comparing time delays in processing of various forms of information.

For example, take the case of implementation of RSA (Rivest, Shamir, Adleman) public key encryption algorithm in a hardware security module. An attacker can encrypt a few thousand plain text samples and note the time it takes each time. With the analysis of this timing information, the attacker can infer the private key stored in the hardware module. Schmeh (2003) proclaims that in the case of a smart card, only a few hours are needed to extract the key.

Timing analysis was first of the side channel attacks and was developed by Kocher. In his article (Kocher, 1996), he has described timing analysis attacks on implementations of Diffie-Hellman, RSA, DSS (Digital Signature Standard) and other crypto systems. By modeling the attack as a signal detection problem, he is able to show how computationally inexpensive the attack can be. The attack is dependent on the fact that digital systems require different amounts of time to process different inputs and similarly the times vary for different types of steps in a given program. For example, the steps involved in branching and conditional statements and processor instructions for multiplication, division or shift operations each require different amounts of time. By making timing measurements, the attacker can infer the step being executed and also the type of data being processed.

To further elaborate the timing attack, consider the RSA decryption operations to compute a plaintext message *m* from cipher text *c* using private key (*d,n*) where *d* is the secret exponent and *n* is the modulus. The computation required to extract message *m* is given below:

$$m = c^d \bmod n$$

An attacker can get samples of cipher text *c* by passively eavesdropping on a target system and *n* can be inferred from the public key (*e,n*). By making timing measurements on multiple decrypted computations of the form given above, the perpetrator of the attack can then infer the secret exponent *d*, thereby enabling him or her to find the secret key (*d,n*). For timing analysis to work, the attacker must also know the encryption algorithm being used by the victim.

**ERROR ANALYSIS**

Error analysis attack is also referred to as fault analysis attack. It was pioneered by Boneh, DeMillo and Lipton (1997) and later developed by Biham and Shamir (1997). In an error analysis side channel attack, the hardware module under attack is falsely activated or damaged and output pattern for a given input is obtained. For example, a smart card is damaged either mechanically or through heat. Output for the same input from a healthy module is also obtained. By comparing the correct and false results, the private key can be reconstructed.

Boneh et al. developed mathematical model of the fault analysis attack based upon transient faults i.e. faults which occur only for a short duration of time and then are gone. For example, flipping of a single bit in a hardware module for a few micro seconds is an example of transient fault. The attacker can also induce the transient faults into a system. Effectiveness of the attack is dependent upon the implementation of the crypto system. For example, for an RSA implementation based upon Chinese remainder theorem, Boneh et al. showed that the modulus can be factored with a very high probability by using a single faulty RSA signature.

The initial fault analysis attack developed by Boneh et al. works against public key crypto systems only and is not feasible against secret key encryption schemes. The attack is based upon algebraic properties of modular arithmetic used in public key encryption algorithms and therefore does not work against secret key encryption algorithms which use bit manipulations instead of arithmetic operations to perform encryption and decryption. Biham and Shamir taking the work further developed differential fault analysis attack which works against both public and secret key encryption schemes. They implemented the attack against an implementation of Data Encryption Standard (DES) and demonstrated that they can extract the DES secret key stored inside a tamper resistant DES hardware encryption device with 50 to 200 known cipher text samples. Even if the DES is replaced by 3DES, the attack can still extract the secret key with the same number of cipher text samples. In the case of encryption algorithms that compute S-boxes as a function of the key, the S-boxes can themselves be extracted.

In addition to these attacks which were developed a few years ago, Takahashi and Fukunaga (2010) demonstrated a successful attack against Advanced Encryption Standard (AES) with 192 and 256-bit keys using C code and a simple personal computer as recent as January 2010. They were able to successfully recover the original 192-bit key using 3 pairs of correct and faulty cipher texts within 5 minutes, and 256-bit key using 2 pairs of correct and faulty cipher texts and 2 pairs of correct and faulty plaintexts within 10 minutes.

## POWER ANALYSIS

In this type of side channel attack, the attacker feeds different inputs to the embedded system and then observes the power consumed. The attacker then draws conclusions about the stored information by measuring and comparing fluctuations in power consumption. For example, DES key embedded in hardware security module can be inferred after about 100,000 encryption operations (Schmeh, 2003).

Like other side channel attacks, the power analysis attack can either be a simple power analysis attack or differential power analysis (DPA) attack. In the case of simple power analysis, the attacker draws conclusion about the operation being performed by observing the amount of power being consumed. For example, different instructions of a microprocessor take different amounts of time to execute and hence consume different amounts of power while they are being executed. Similarly, different stages of an encryption algorithm take different amount of time and power to execute. Some stages may be more computationally extensive and hence require more amount of power to execute and some other stages may require less amount of power to execute. As a result, by observing the power being consumed at a particular instant of time, the attacker can infer information about the stage of the encryption algorithm being executed and also the data upon which the operation is being performed. In other words, a simple power analysis can reveal the sequence of instructions being executed by the system under attack. Cryptosystems in which the path of execution depends upon the type of data being processed can be broken using this knowledge gained about the sequence of instructions being executed.

In the case of differential power analysis, the attacker not only makes observations of power consumed but also a statistical analysis of the data is carried out to infer the data value and the operation being performed on it. As stated earlier, there can be power variations due to various stages of an encryption algorithm being executed. In addition to this, there will also be power variations due to data values being operated upon in a particular stage. However, power variations due to later are much smaller than those due to the former reasons. These smaller variations are usually lost due to error measurements and other reasons. In these cases, a statistical power model, particularly developed for a target algorithm under test, correlating power consumption due to various stages and different types of data values can be applied to infer the secret values. In the first stage, observations on the power being consumed are made and in the second stage this data is fed to a statistical model. Error correction procedures are also applied to gain more accurate picture of the operations being performed by the system under attack. Typically a cryptographic algorithm uses only parts of the secret key, called sub keys, at certain stages. Instead of running the full blown algorithm, the attacker can write program to simulate selected part of the algorithm where computations involving sub keys are made. The attacker then calculates the intermediate values of that particular stage for all possible sub key guesses. The calculated intermediate values are fed to the statistical power model to predict the power consumption for that computation and hence the corresponding sub key. Then the attacker runs the real cryptosystem under attack with the same input data and makes an observation of the power consumed. The observed power consumption value is compared with the values obtained from the statistical model. All the power consumption values obtained from the statistical model that do not match with the real power consumption value are derived from wrong key guesses. However, matching power consumption value from the statistical model is derived from the correct sub key guess. As a result, the attacker is able to isolate correct sub key guesses from the wrong ones. In this way, by comparing the real power consumption values with those obtained from the statistical model, the secret key can be inferred.

Differential power analysis attack was initially suggested by Kocher, Jaffe and Jun (1999). Later, Messerges, Dabbish and Sloan (2002) extended the research and provided experimental data and attack details on a smart card. Brier, Clavier and Oliver (2004) further investigated the DPA attack and used classical model for the power consumption of cryptographic devices. The model is based on the Hamming distance of the data handled with regard to an unknown but constant reference state. Once the model is validated, it allows to mount an optimal attack called Correlation Power Analysis (CPA). This attack is similar to the one described earlier except the fact that the power model used as a reference is different from the statistical model.

In addition to the above attacks on crypto systems based on smart cards, the DPA attacks have been shown to be successful on ASIC (Application Specific Integrated Circuit) and FPGA implementations of various encryption algorithms. In particular, Standaert, Ors, Quisquater and Prencel (2004) demonstrated a successful attack against an FPGA implementation of the DES and Ors, Gurkaynak, Oswald and Prencel (2004) demonstrated a successful attack against an ASIC implementation of the AES. Lately, AES encryption algorithm and its implementations have been subjected to various types of tests and attacks. Han, Zou, Liu and Chen (2008) and Kamoun, Bossuet and Ghazel (2009) have independently developed experimental attacks against various types of hardware implementations of this algorithm.

Differential power analysis is one of the most popular side channel attacks because it is easier to conduct and can be repeated without damaging the object under analysis. In particular, smart cards, which typically consist of an 8-bit processor, EEPROM, small amount of RAM and a small operating system have been a particular target of DPA attacks.

**ELECTROMAGNETIC RADIATION ANALYSIS**

Even if an embedded system does not house a Trojan horse, or is not prone to timing, power or error analysis, it is still possible to breach its security by other means. In an attack based on electromagnetic radiation (EMR) analysis, the attacker captures and reconstructs the signal leaked through the electromagnetic radiation from the target equipment.

It is well known that the US government has been well aware of attacks based on analysis of electromagnetic radiation since 1950s and that display screen of video display units could be reconstructed after capturing their EMR. Standards were developed for the protection against this attack and were called TEMPEST which is an acronym for Telecommunications Electronics Material Protected from Emanating Spurious Transmissions. Partial details of TEMPEST are also available at the Internet. TEMPEST certification for private organizations is very expensive and therefore another standard called ZONE has been developed which is less secure but costs less than TEMPEST. There are three classes of TEMPEST standard: class 1 has the highest level of security and is available only to US government and approved contractors; class 2 is less secure and is again meant for use of US government; class 3 is available for general and commercial purposes.

In his landmark paper, Van Eck (1985) showed that display screen of even standalone systems could be reconstructed within a distance of 2 km by collecting and processing electromagnetic radiation of the display. Van Eck was able to capture the display screen of a computer using a normal TV receiver and a piece of small extension equipment costing only US$ 15.

EMR analysis attack is particularly dangerous against digital signals. If the digital signals of data being processed by a device can be reconstructed remotely, this can reveal the data. For example, hard disk stores information in binary form and when the data is read from or written to the hard disk, the digital signals are generated during these operations. If these signals are strong enough to be reconstructed remotely, the data being read from or written to the hard disk can be seen by the attacker.

Wright (1987), who was a senior intelligence officer in the British Intelligence MI5, revealed in his book, Spycatcher – The Candid Autobiography of a Senior Intelligence Officer, how he was able to reconstruct the messages sent by the French diplomats. At first, he tried to break the cipher but failed. However, Wright and his assistant noticed that the encrypted traffic carried a faint secondary signal. Further analysis revealed that it was the electromagnetic radiation of the encryption machine and signal reconstruction provided them the plain text without having to break the cipher.

EMR analysis attack has also been shown to work against hardware implementations of encryption algorithms. De Mulder et al. (2005) have demonstrated that a simple EMR analysis attack against elliptic curve cryptosystem is able to find all the key bits with a single measurement. Similarly, a differential EMR analysis attack against an improved implementation of the same cryptosystem requires approximately 1000 measurements to find the key bits.

In addition to the four different types of side channel attacks discussed above, there may be attacks which are indirect attacks on embedded systems security. For example, networked embedded systems typically use TCP/IP suite of protocols to communicate with each other and a central processor. There are many proven flaws in TCP/IP suite of protocols which are inherent to the design of these protocols. In addition, there are other vulnerabilities in the implementation of these protocols which an attacker can target to breach the security of an embedded system. All these flaws can be used to exploit and bug the networked embedded systems themselves or communication amongst them.

**COUNTERMEASURES AGAINST SIDE CHANNEL ATTACKS**

Development of attacks against hardware and embedded systems has prompted researchers to design appropriate countermeasures against each of these types of attacks.

A countermeasure suggested against timing and power analysis is the insertion of random delays at various stages of the data processing. The delay insertion may be possible at any layer of abstraction. After the delays have been added in a random manner, it will become difficult for the attacker to guess the nature of operations being performed or the data being processed. Similarly, the introduction of delay will also result in change of power profile of the device increasing the complexity of power analysis attack. It may be noted that adding a number of delays will also result in slowing down of the crypto system. Further, adding delay will not significantly reduce the probability of a successful attack and will only increase the complexity of a feasible attack on the system. For example, the attacker will now have to make some more measurements and arrive at some more precise modeling to infer the secret information stored in the hardware device.

It is intuitive to think that an improved and better prevention mechanism against timing analysis would be to make all computation operations to take the same amount of time. However, it is difficult to implement and will render the system too slow to be usable. Instead, a better alternative would be to take out frequently used operations, say, multiplication, division, exponentiation, and then make them execute in a fixed time. In other words, instead of having all instructions to execute in a fixed time, we take out only the frequently used instructions and then make them execute in fixed time. In this way, majority of the operations in the digital system are executed in a fixed time and this significantly increases the complexity of guessing the operation being performed or inference of information stored in the digital system.

A countermeasure against the fault analysis attack would be to run the encryption algorithm twice and then compare the results. The computation is considered valid only if the two results match (Potkonjak, Nahapetian, Nelson, & Massey, 2009). However, this will significantly increase the computation time of the algorithm. Further, the fault analysis attack is still not impossible to mount and only the number of computations required by the attack increase so as to increase the level of complexity of the attack.

As power analysis attacks are more common, there are a number of countermeasures suggested against these types of attacks. For example, Tiri and Verbauwhede (2004) have suggested using Simple Dynamic Differential Logic (SDDL) and Wave Dynamic Differential Logic (WDDL) to build basic gates all of which consume the same amount of power. Any logic implemented with these gates will then have constant power consumption. Similarly, dummy gates and logic can be inserted in a circuit so that the power consumed becomes equal in all operations. Further, by avoiding branching, conditional jumps, etc. many power analytic characteristics of the system can be masked. Another countermeasure suggested against power analysis attacks is the addition of random calculations so as to add dummy peaks in the power model constructed by the attacker.

Primary means of protection against EMR analysis is the use of shields (Faraday cages) in the equipment which is to be protected. In addition to this, the equipment which is to be protected against EMR analysis uses components that have low EMR.

## SECURITY OF CELL PHONES

Cell phones, which are a particular category of embedded systems, have specially been a target of various types of attacks ranging from malformed SMS (Short Message Service) text messages and bluetooth packets to cloned SIMs. Our discussion of embedded systems security will not be complete without mentioning issues related to the security of cell phones.

Today's cell phones and PDAs are complex and complicated embedded systems with powerful options. If someone can remotely take control of a cell phone, it could, for example, send SMS messages to the contacts stored in the cell phone or other numbers without knowledge of the owner. It could also send a busy tone to a caller when the phone is in fact not busy or it could be calling the numbers at random. Even it can turn on the microphone in the cell phone and eavesdrop on any conversation going on in the vicinity of the cell phone. In fact, it has been revealed that a cell phone can be used to eavesdrop on conversation even when it is turned off as powering off may only cut off power to the LCD display while the internal circuitry remains operational. McCullagh and Broache (2006) disclosed that FBI has been using cell phones in this way for surveillance of criminals.

Attacks on cell phones can usually be mounted in any of the following ways:
- Attacks using the bluetooth communication protocols
- Attacks using text messages
- Attacks on the SIM card for GSM and CDMA cell phones

Bluetooth is a wireless communication protocol operating at 2.4 GHz for short range communications and one of its objectives is to replace cables and wires connecting peripheral devices. Cell phones and PDAs have a bluetooth interface for a remote headset and also to facilitate communication with other cell phones. However, this also makes it possible to mount remote attacks on the cell phone by exploiting weaknesses in the bluetooth protocol.

First step in an attack against a bluetooth device is building of a malformed object i.e. a legitimate file with invalid contents arranged in such a manner to break the target system. The object is then sent to the victim causing the receiving device to behave as programmed in the malformed object which is usually to allow the attacker to take control of the device.

There are many forms of attack that can be launched using the bluetooth protocol thereby giving rise to a plethora of jargon like bluejacking, bluesnarfing, bluebugging, bluedumping, bluesmack, bluespoofing, etc. Bluejacking is a technique whereby the attacker sends a malformed vCard with message in the name field and exploits OBject EXchange (OBEX) protocol. Bluesnarfing works by using the OBEX push profile to connect to bluetooth device which is in discovery mode. With this attack, the attacker can extract important information from the cell phone e.g. phonebook. Bluebugging is a form of attack that exploits the victim's AT command parser to allowing the attacker to take complete control of victim's cell phone. After taking control, the attacker can virtually make any use of victim cell phone like placing of calls, sending and receiving of text messages, change phone and service settings. Both free and commercial tools are available to exploit the bluetooth protocol. These include Blooover [sic], BTCrack, Bluesnarfer, BTClass, carwhisper, BT Audit, Blueprint, etc.

In an attack exploiting vulnerability in the SMS text message module of the cell phone, the attacker crafts a special SMS message which the victim cell phone is unable to handle so as to cause a crash or cause the execution of payload so as to return control of the victim phone to the attacker. The attacker must know the particular layout of the SMS that will cause the desired behavior on the victim cell phone. To find out such an SMS layout, the attacker may test the particular model of the victim cell phone by fuzzing. Mulliner and Miller (2009) presented a fuzzing methodology for smart cell phones in an information security event. They proposed techniques which would allow a researcher to inject SMS messages into smart phones without using the carrier i.e. the phone service provider. They also demonstrated how they were able to discover a flaw in Apple's iPhone so that it could be hacked using a single SMS message.

Another possible attack on GSM (Global System for Mobile Communication) and CDMA (Code Division Multiple Access) cell phones is the cloning of the Subscriber Identification Module i.e. the SIM card. A SIM card is smart card which contains information, such as the serial number, International Mobile Subscriber Information (IMSI), the cryptographic algorithms A3 and A8, secret key Ki, two passwords i.e. PIN which is for usual use and PUK (PIN Unlock Key) for unlocking the phone and Kc which is derived from Ki during the encryption process. If an attacker gets hold of a victim's SIM card and is able to read the data on the SIM card using a SIM card reader, then she can prepare a duplicate (clone) of the SIM using a SIM card writer. The cell phone holding the cloned SIM can authenticate and communicate with the cell phone service like the original cell phone. However, the attack is not as easy as the above description has presented it. The real problem faced by the attacker is the extraction of secret information from the SIM which is tamper resistant. The GSM cell phones use an algorithm called COMP128 for authentication and derivation of keys. This algorithm was used only to serve as an example by the GSM standard. The mobile operators were supposed to choose a better algorithm replacing COMP128. However, majority of mobile operators went along using COMP128. Later on many weaknesses were discovered in the COMP128 algorithm which allowed an attacker having physical access to a SIM card to exploit the weakness in COMP128 and retrieve the information stored inside a SIM card. This allowed the attacker to prepare a duplicate SIM using the SIM card writer as described above. This remained a major problem for some time until new revisions of COMP128 algorithm were developed. The initial version of COMP128 is now referred to as COMP128-I and subsequent versions are referred to as COMP128-2, COMP128-3 and COMP128-4. Later versions are secure and there are no known attacks against them as yet. As a result, copying of information from a SIM card is only possible in case of old SIM cards employing COMP128-1. Newer SIMs employing later versions of the algorithm are not susceptible to any known attacks and do not reveal stored information and hence do not permit their duplication.

**FUTURE RESEARCH DIRECTIONS**

As embedded systems security is an emerging area of research, the hardware security research community is expected to discover and develop new forms of attacks on the hardware devices and embedded systems. This will further fuel the research on the countermeasures and protection schemes against these attacks. As a result, there may be a paradigm shift in the design of hardware devices and embedded systems particularly those used in defense applications. For example, design and implementation of an IC may go through fundamental changes from an abstract level description to semiconductor level fabrication so as to incorporate new security measures. For this to happen, the research community needs to come up with reliable and robust techniques which can be implemented at every layer of abstraction.

Embedded systems security shares a few common traits with software security. Therefore, in these common areas existing security techniques and methods may be applied. However, a few of the traits of embedded systems security are different from those of software security. Therefore, new algorithms and techniques will need to be developed for these areas. For example, research community will need to build secure operating systems which can be deployed in embedded systems which are usually resource constrained. This will include implementing security in all the critical functions performed by an operating system.

In many instances, attacks on embedded devices are possible only if physical security of the device is compromised. For example, in the case of side channel attacks on smart card, the attacker first needs to get a copy of the smart card which she intends to attack. Therefore, new security techniques will need to be developed which can prevent physical tampering of the device and ensure that the information stored in it remains secure.

Embedded devices are usually limiting in resources and most of the existing cryptographic algorithms are computation intensive. Implementation of security with the heavy algorithms results in performance degradation. Therefore, lightweight cryptographic algorithms and protocols are needed which are tailored to run on embedded devices with limited resources.

# CONCLUSION

Embedded systems are finding widespread uses in every sphere of our lives and their security has become an important research issue. In this chapter, we have discussed the background and current state of research on the threats and attacks being developed against embedded systems. The hardware attacks can be mounted at any of the layers of abstraction involved in the fabrication of the device with varying degrees of success. We have also discussed various countermeasures against these attacks.

**ADDITIONAL READING**

## KEY TERMS AND DEFINITIONS

**Covert channel:** is a hidden communication channel between a malicious entity and a victim system which is conventionally not used for transfer of information.

**Electrically Erasable Programmable Read-Only Memory (EEPROM):** is a type of Read Only Memory (ROM) which is used in embedded systems and computers to store data or boot program which must not be destroyed when power is turned off. The data or program can only be erased or programmed electrically.

**Embedded system:** is a digital system which is designed to perform a specific task or set of tasks as compared to a general purpose computer which allows user to perform various types of tasks by installing new software on top of its operating system.

**Field Programmable Gate Array (FPGA):** is a type of integrated circuit which can be programmed by the design engineer to implement a particular logic at the hardware level. An HDL is used to program an FPGA.

**Finite state machine (FSM):** is an abstract mathematical and behavioral model of a system consisting of a number of finite states that the system may switch to and a finite number of inputs that cause the transition from one state to the other. An FSM is also commonly known as finite state automation or simply state machine.

**Fuzz Test:** or fuzzing is a testing technique in which the system is provided with various combinations of invalid input and behavior of the system is observed.

**Hardware Description Language (HDL):** is a computer programming language which is used to describe the logic and flow of an electronic circuit. Examples of HDL are VHDL and Verilog.

**Hardware Trojan horse:** is a bug or backdoor in the form of a tangible circuit or software piece of code in the control program of an electronic system. It may allow an unauthorized user to communicate with the system and control it.

**S-Box:** is an acronym for Substitution-Box and is a type of lookup table which is used in a symmetric key encryption algorithm to perform substitution operation on a given plain text. The box receives *m* number of input bits and according to some lookup function, translates them to an *n* number of output bits.

**Side channel attack:** is a type of attack on an embedded system which treats the embedded system as a black box and tries to infer hidden information by observing the behavior (timing, power consumed, etc.) and output of the system by feeding it various types of inputs.

**Smart card:** is a small pocket-sized plastic card which has an integrated circuit (usually an embedded processor) inside it.

**Subscriber identification module (SIM):** is a smart card used in GSM and CDMA based cellular telephones to allow users to switch phones by simply switching the SIM.